\documentstyle[aps,prl,multicol,epsf]{revtex}
\begin{document}
\draft
\title{
Reaction-Diffusion-Branching Models of Stock Price Fluctuations
}
\author{Lei-Han Tang$^{(1)}$ and Guang-Shan Tian$^{(1,2)}$}
\address{
$^{(1)}$ Department of Physics, Hong Kong Baptist University,
Kowloon Tong, Hong Kong\\
$^{(2)}$ Department of Physics, Peking University,
Beijing 100871, China
}

\date{\today}
\maketitle
\begin{abstract}
Several models of stock trading [P. Bak {\it et al},
Physica A {\bf 246}, 430 (1997)] are analyzed in analogy with
one-dimensional, two-species reaction-diffusion-branching processes.
Using heuristic and scaling arguments, we show that the short-time
market price variation is subdiffusive with a Hurst exponent
$H=1/4$. Biased diffusion towards the market price and blind-eyed copying
lead to crossovers to the empirically observed random-walk 
behavior ($H=1/2$) at long times. The calculated crossover forms and 
diffusion constants are shown to agree well with simulation data.

\end{abstract}

\begin{multicols}{2}

The movement of stock prices is among the oldest class of
fluctuation phenomena that have been analyzed quantitatively
\cite{history,mandelbrot}, yet very limited understanding on
the origin and size of market volatilities --- so fundamental to much of
the finance literature\cite{hull} --- is available to date.
Recent empirical studies\cite{mandelbrot,stan,bouchaud} of major 
financial indices around the world have revealed a number of
universal features in the time series, including Levy-like
distribution of short-time price moves. In search of 
unifying principles that explain the observed behavior,
Bak, Paczuski, and Shubik (BPS)\cite{bps} recently proposed
several models of stock trading among interacting agents.
Their numerical study of these models suggests a rich set
of dynamical behavior, including stock price fluctuations
which exhibit similar statistical patterns as those of real 
markets\cite{bouchaud}.

From a statistical mechanical point of view, the family of models
proposed by BPS are particularly interesting due to their connection
to reaction-diffusion processes familiar in physical 
contexts\cite{ovch}, and thus
one may hope to gain some insight into the
collective behavior of traders by exploiting this analogy.
A mapping between the two is easily constructed by taking 
the target price of agents as their 
coordinates on a one-dimensional price axis.
The two types of agents, i.e., buyers and sellers of a stock,
are identified as two species, $A$ and $B$, respectively.
At any given moment, the population of buyers are separated
from that of sellers by the market price $x_M(t)$ where
transactions, or for that matter ``reactions''
$A+B\rightarrow\emptyset$, take place.
Such reaction-diffusion problems have been studied
extensively in the past\cite{redner,cornell,cardy,krap}.
A result of particular interest is 
the power-law scaling of the reaction front fluctuations,
\begin{equation}
\langle [x_M(t)-x_M(t')]^2\rangle^{1/2}\sim |t-t'|^H,
\label{scaling}
\end{equation}
where the exponent $H=1/4$ (with possible logarithmic correction).
There are, however, two new elements in the BPS models which
have not been examined before: biased diffusion 
of $A$ and $B$ particles towards the reaction front, and
price ``copying'' which translates to branching
$A\rightarrow 2A$ and $B\rightarrow 2B$.
BPS showed numerically that, in the latter case, the long time
behavior of $x_M(t)$ changes to that of a random walk with $H=1/2$.

The purpose of this paper is to establish an analytic foundation
for various observations made by BPS in their pioneering work
and also to further quantify and extend their numerical results.
We identify the driving force of the market price variation
and determine the size of the market response from the distribution
of agents near the market price.
The analysis yields not only the scaling exponent $H$, but
also the scaling amplitudes and various crossovers. 
Good agreement is reached
between theoretical predictions and simulation data for
a broad range of model parameters.

The original BPS model is a trading game with equal
number ($N/2$) of buyers and sellers, each attaches a
price $x_i$ to the stock they intend to buy or sell.
A buyer owns no share and a seller owns exactly one share.
The agents perform simultaneous, independent random walks on the price
axis until they meet a member of the other group. Upon transaction,
buyer and seller exchange their role and are then relocated 
on the price axis. 
In this paper we shall consider
three variants of the model as detailed below:

{\it Model I (unbiased diffusion)} --- The price $x_i$ of agent $i$
moves up or down by one unit with equal probability in each time step.
For convenience, we decouple the
reaction event $A+B\rightarrow\emptyset$ from the relocation of the 
agents in a transaction (see note \cite{note1}). 
A steady-state situation is maintained by
injecting new agents from the two ends of a prescribed price interval
at a given rate $J$.

{\it Model II (biased diffusion)} ---
The rules in this case are similar to those of Model I except that the 
updating of $x_i$ is biased towards the market price. Specifically, for
a buyer, $x_i\rightarrow x_i+1$ with probability $(1+D)/2$ and
$x_i\rightarrow x_i-1$ with probability $(1-D)/2$. The rule is reversed
in the case of sellers. Obviously, Model I is regained by setting
$D=0$.

{\it Model III (biased diffusion with copying)} ---
The updating of $x_i$ is the same as in Model II but now, after
a transaction, the buyer and seller are immediately 
re-injected into the market by duplicating the price of a fellow agent
chosen at random. 
According to BPS\cite{bps}, such a process imitates 
herding behavior in real markets. 
In the particle language, it can be represented by 
stochastic branching $A\rightarrow 2A$ and $B\rightarrow 2B$.

We start our discussion by considering a modification
of the above models which is minor from the point
of view of a given diffusing particle in its whole lifetime
(i.e., from its first release to the reaction),
but it trivializes the problem completely.
Instead of asking an $A$ particle to find a $B$ particle
for a reaction, we assume that the reaction always takes
place at a fixed position, say $x=0$. In essence, 
we are making the assumption that the market price fluctuates
at a much slower rate compared to the diffusive motion
of individual agents, a commonly used approximation 
in the study of interface fluctuations\cite{cahn}.
The point $x=0$ now serves as a trap of the diffusing particles
which do not interact with each other.
In the continuum limit, the average density of buyers $a(x,t)$ 
and sellers $b(x,t)$ obey the following linear equations,
\begin{mathletters}
\begin{eqnarray}
\partial_t a &=& \gamma\partial_x^2 a - \beta \partial_x a+S_A,
\label{den_a}\\
\partial_t b &=& \gamma\partial_x^2 b + \beta \partial_x b+S_B,
\label{den_b}
\end{eqnarray}
\label{density}
\end{mathletters}
with the boundary condition $a(0,t)=b(0,t)=0$.
Here $\gamma$ is the diffusion coefficient of individual particles
and $\beta$ describes drift towards the current market price. 
The updating rule of $x_i$ given above 
specifies $\beta=D$ and $\gamma=(1-D^2)/2$ (parallel updating)
or $\gamma=1/2$ (random sequential updating).
The source terms $S_A$ and $S_B$ correspond to injection
of new particles into the system. 
We now determine the steady-state solutions
$a_0(x)$ and $b_0(x)=a_0(-x)$ to Eqs. (\ref{density}).

{\it Models I and II}. --- The particle current
\begin{equation}
J=-\gamma\partial_x a_0+\beta a_0,
\label{J_scheme_I}
\end{equation}
is a constant in this case.
For $\beta=0$ (Model I), the current is maintained by a
linear profile [Fig. 1(a)], 
\begin{equation}
a_0(x)=-Jx/\gamma.
\label{a_0_I}
\end{equation}
For $\beta>0$ (Model II), $a_0(x)$ crosses over from the linear function
(\ref{a_0_I}) close to the origin to a constant $J/\beta$ at large 
distances [Fig. 1(b)],
\begin{equation}
a_0(x)=(J/\beta)[1-\exp(\beta x/\gamma)].
\label{a_0_II}
\end{equation}

{\it Model III}. ---
Branching introduces
source terms $S_A=\alpha a$ and $S_B=\alpha b$, where $\alpha$
is a branching rate.
The equation for $a_0(x)$ is now a second-order ordinary
differential equation which can be solved to yield,
\begin{equation}
a_0(x)=C[\exp(k_-x)-\exp(k_+x)],
\label{a_0_III}
\end{equation}
where $k_\pm=(\beta\pm\sqrt{\beta^2-4\alpha\gamma})/(2\gamma)$
and $C$ is an overall amplitude.
The steady-state solution exists only when
$\alpha\leq\alpha_c=\beta^2/4\gamma$.
The shape of the profile is indicated in Fig. 1(c), which
is linear close to the origin and decays exponentially at 
large distances. The current of incoming particles at the origin is
given by $J=-\gamma\partial_x a_0=\gamma C(k_+-k_-)$.

\begin{figure}
\narrowtext
\epsfxsize=\linewidth
\epsffile{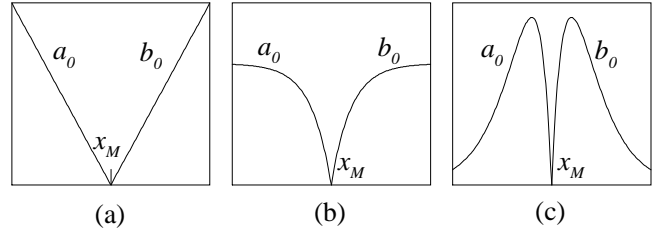}
\caption{Average density profiles of buyers ($a_0$) and sellers ($b_0$)
along the price axis for
(a) Model I (unbiased diffusion); (b) Model II (biased diffusion); 
(c) Model III (biased diffusion with copying).
Note that in all three cases $a_0(x)$ and $b_0(x)$ vanish linearly
at the market price $x_M$.
}
\label{fig1}
\end{figure}

The total number of, say $A$ particles $n_A$ that arrive at the trap 
in a time interval $t$ to $t'=t+\tau$ is a fluctuating quantity 
which can be expressed as,
\begin{equation}
n_A=\sum_i \eta_i.
\label{N-walker}
\end{equation}
Here the sum runs over all particles $i$ that entered the 
system from the beginning of the process to time $t'$,
and $\eta_i$ is a random variable
which takes the value one if particle $i$ is trapped during the interval
$\tau$ and zero otherwise.
Since the $\eta_i$'s are independent from each other\cite{note2}, we
easily find,
\begin{mathletters}
\begin{eqnarray}
\langle n_A\rangle &=&\sum_i p_i,
\label{mean}\\
\langle n_A^2\rangle-\langle n_A\rangle^2
&=&\sum_i (p_i-p_i^2),
\label{sigma}
\end{eqnarray}
\end{mathletters}
where $p_i$ is the probability that $\eta_i=1$.
For $p_i\ll 1$, which holds when $\tau$ is much smaller than
the typical spread of the lifetime of the diffusing particles,
we have the following approximate relation,
\begin{equation}
\langle n_A^2\rangle-\langle n_A\rangle^2\simeq\langle n_A\rangle
=J\tau,
\label{n_A_fluc}
\end{equation}
where, as before, $J$ is the flux of particles entering the trap.
Results derived below are based on this approximation but
other situations may also be considered.

We now construct a heuristic argument to show how the fluctuations
in $n_A$ and $n_B$ lead to a shift of the reaction front or the market price.
To be definite, we take $\Delta n=n_A-n_B>0$ so that an upward move
of the market price from $x_M$ at time $t$ to $x_M'$ at time
$t'=t+\tau$ is expected (see Fig. 2).
Loosely speaking, the interval $[x_M,x_M']$ defines a
reaction zone within which most of the $n_A$ particles entered
through $x_M$ reacted with the $n_B$ particles
entered through $x_M'$. 
The excess number of $A$ particles (buyers) $\Delta n$
have either reacted with the $B$ particles initially in the
zone at time $t$, or remained in the zone at the end of the period.
Based on this observation, we may identify $\Delta n$ with the
sum of shaded areas under $b(x,t)$ and $a(x,t')$ in Fig. 2,
respectively,
\begin{eqnarray}
\Delta n&\simeq& 
\int_{x_M}^{x_M'} a(x,t')dx+
\int_{x_M}^{x_M'} b(x,t)dx\nonumber\\
&\simeq&2\int_0^{\Delta x}a_0(x-\Delta x)dx.
\label{fund_rel}
\end{eqnarray}
Here $\Delta x=x_M'-x_M$ is the price move over the time period $\tau$.
A similar relation holds for $\Delta n<0$.

\begin{figure}
\narrowtext
\epsfxsize=\linewidth
\epsffile{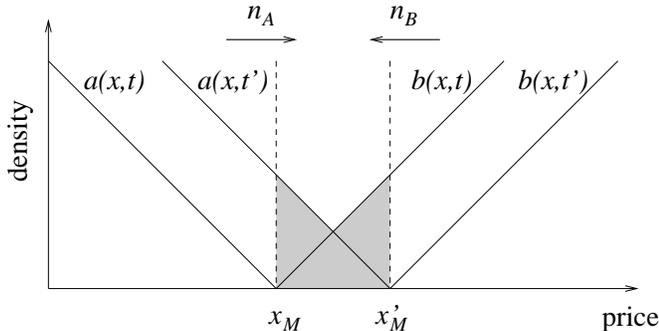}
\caption{Schematic illustration of how an excess number of particles
$\Delta n=n_A-n_B$ reaching the reaction zone $[x_M,x_M']$ 
is accommodated by a shift of the density profiles 
at time $t$ to those at a later time $t'$.
Graphically, $\Delta n$ is identified with the sum of the shaded areas 
under $a(x,t')$ and $b(x,t)$, respectively.
}
\label{fig2}
\end{figure}

Equation (\ref{fund_rel}), which holds in an average sense,
is our fundamental relation that
links the price move $\Delta x$ to the ``demand-supply imbalance''
$\Delta n$ through the density profile $a_0(x)$.
From the statistics of $\Delta n$ we can then work out the
statistics of $\Delta x$.
Below we give our results on the market price fluctuations
using this approach.

{\it Model I}. ---
In this case, the intrinsic profile $a_0(x)$ 
is linear. From Eqs. (\ref{a_0_I}) and (\ref{fund_rel}),
we obtain $\Delta n\simeq J(\Delta x)^2/\gamma$.
From the variance $\langle\Delta n^2\rangle=2J\tau$, we obtain,
\begin{equation}
\langle [x_M(t)-x_M(t')]^2\rangle^{1/2}\simeq 
\bigl({4\gamma^2\over\pi J}\bigr)^{1/4}|t-t'|^{1/4}.
\label{I_0_scaling}
\end{equation}
This form agrees with previous results of 
Refs. \cite{cornell,cardy,krap}.
(At very short times, the discreteness of $\Delta n$
manifests itself which leads to deviations
from Eq. (\ref{I_0_scaling}). See also Ref. \cite{redner}.)

{\it Model II}. ---
Since $a_0(x)$ goes to a constant for $|x|>x_c=\gamma/\beta$,
Eq. (\ref{fund_rel}) yields a linear dependence
$\Delta n\simeq 2J\Delta x/\beta$ for $\Delta x>x_c$.
Hence the move of $x_M(t)$ at long times is a random walk
with a diffusion constant 
$\Gamma\simeq \langle\Delta x^2\rangle/2\tau=\beta^2/(4J)$.
More detailed calculation yields a crossover scaling,
\begin{equation}
\langle [x_M(t)-x_M(t')]^2\rangle^{1/2}\simeq 
x_c\Phi\Bigl({|t-t'|\over t_c}\Bigr),
\label{I_1_scaling}
\end{equation}
where $t_c= \gamma^2J/\beta^4$. The limiting forms of the scaling function
are given by $\Phi(s)\simeq (4s/\pi)^{1/4}$ for $s\ll 1$ and 
$\Phi(s)\simeq (s/2)^{1/2}$ for $s\gg 1$. 

{\it Model III}. ---
In this case the profile (\ref{a_0_III}) extends only
over a finite range of $x$, so the finite lifetime of a particle
becomes an important factor in our consideration.
The short time behavior of the price fluctuation
is similar to that of model I and II due to the linear
behavior of $a_0(x)$ close to the origin,
which is common in all three cases.
Thus Eq. (\ref{I_0_scaling}) can still be applied in this regime.
Crossover to a different behavior is expected when
$\tau=t'-t$ becomes comparable to the lifetime of
a particle $\tau_0=2\gamma/\beta^2$. In fact,
$\tau_0$ is also the relaxation time of the density
profiles as can be seen by bringing Eq. (\ref{density})
into a dimensionless form. 
On time intervals larger than $\tau_0$, memory about the
initial profile is essentially lost and the next move of the
market price is equally likely to be up or down, hence
a random walk behavior with a step size set by the size of
the fluctuation $x_0=\gamma^{1/2}J^{-1/4}\tau_0^{1/4}$ at $\tau=\tau_0$
[see Eq. (\ref{I_0_scaling})]. 
The usual scaling argument then yields,
\begin{equation}
\langle [x_M(t)-x_M(t')]^2\rangle^{1/2}\simeq 
x_0\Psi\Bigl({|t-t'|\over\tau_0}\Bigr),
\label{I_2_scaling}
\end{equation}
where $\Psi(s)\simeq s^{1/4}$ for $s\ll 1$ and
$\Psi(s)\sim s^{1/2}$ for $s\gg 1$.
The diffusion constant of the market price at long times is
given by $\Gamma'\simeq x_0^2/\tau_0=\beta\gamma^{1/2}J^{-1/2}$.

We have performed numerical simulations of the BPS models
to check the validity of the theoretical analysis presented
above. Since our results on Model I are similar to those of previous
studies (apart from a possible logarithmic correction), we shall focus on
Model II and III. 

Model II was simulated at $J=\beta$ where
the asymptotic density is one as in Ref.\cite{bps}.
The system size is chosen to be $N=2000$
or larger to ensure that the reaction front does not fluctuate
out of the boundaries during the time period simulated. 
Otherwise, $N$ is found not to have any significant effect 
on our results\cite{note1}.
The system is first equilibrated for a period $t_0=10\tau_0$ where
$\tau_0=N/(2\beta)$ is the typical lifetime of a particle.
The market price time-series $x_M(t)$
is then recorded over 500 successive time segments, each of 
length $8192$ time steps. We then calculate
$\langle [x_M(t)-x_M(t')]^2\rangle$ averaged first over each time segment
and then over different segments. In Fig. 3(a) we plot
the simulation results using scaled variables for $D=0.01$ to 0.5.
There is indeed a good data collapse over six decades. In fact,
for $t>t_c$, not only the scaling exponent, but also the scaling
amplitude are borne out by the data. 

The simulation of Model III was carried out in a similar way
as that of Model II, except the number of particles
$N$ is now fixed. 
To compare the simulation data with Eq. (\ref{I_2_scaling}),
we use the relation $J=\alpha N/2$ (particle conservation)
from the solution (\ref{a_0_III}). Taking 
$\alpha=\alpha_c=\beta^2/4\gamma$\cite{note3},
we obtain $x_0=2\gamma\beta^{-1}N^{-1/4}$. 
In Fig. 3(b) we plot
the simulation results for market price fluctuations using the
scaling suggested by Eq. (\ref{I_2_scaling}). For four different
values of $D=\beta$ and two system sizes $N=400$ and 1000,
good data collapse is again achieved.

\begin{figure}
\narrowtext
\epsfxsize=\linewidth
\epsffile{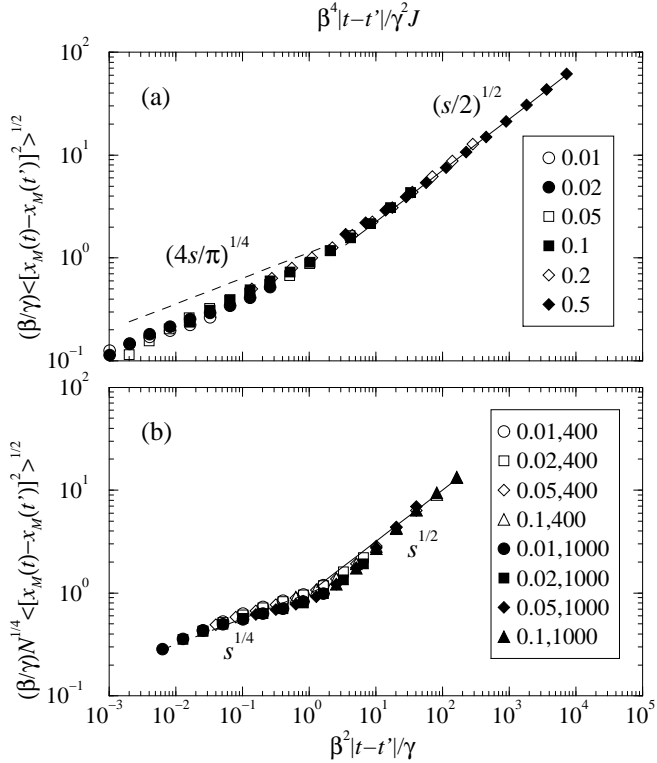}
\caption{Scaling plots of the market price fluctuations versus
time increments for a wide range of model parameters.
(a) Model II. The value of $D$ for each data set is given in the legend.
(b) Model III. The values of $D$ and $N$ for each data set are given
in the legend.
}
\label{fig3}
\end{figure}

In summary, we presented a heuristic method to link the market
price fluctuation to the diffusive motion of individual agents
using the BPS models as examples.
The analysis yields qualitative as well as quantitative
predictions on the size of the market price fluctuations
as a function of time, the number of traders in the market,
and various other model parameters.
For short times, a previously known $H=1/4$ scaling law
is rederived and its validity is correlated to the 
generic linear shape of population density profiles
near the market price. Crossover to the long-time random walk
behavior with $H=1/2$ takes place when agents are driven
to the market price via a diffusion bias. 
Expressions for the crossover time as a function of various
model parameters are derived. These results are shown to
compare favorably with the simulation data.

The $H=1/4$ scaling at short times is quite remarkable and
is against the prevailing thinking in finance that, in a 
market with noise traders only, there should be no restoring
force to price moves and hence no correlation in the market price time series.
Although the exponent is not new, the analysis presented here
makes it plain that resistance to price change is inherent
in the existing price distribution of agents. 
It remains to be elucidated
how such tendencies are modified when external 
information (e.g., financial news) are fed into the market.

One of us (G.S.T) would like to thank
the Croucher Foundation for financial support and the Physics
Department, Hong Kong Baptist University for their hospitality.


\end{multicols}


\begin{references}

\bibitem{history}
L. Bachelier, in {\it The Random Character of Stock Market Prices}, 
edited by P. H. Cootner (MIT Press, 1964), p. 17
(translation of 1900 French edition).

\bibitem{mandelbrot}
B. Mandelbrot, J. Business {\bf 36}, 394 (1963);
E. E. Peters, {\it Fractal Market Analysis} (Wiley, N.Y., 1994).

\bibitem{hull} F. Black and M. Scholes, J. Political Economy
{\bf 81}, 637 (1973); P. Wilmott, S. Howison, and J. Dewynne,
{\it The Mathematics of Financial Derivatives}
(Cambridge,  1995);
K. Cuthbertson, {\it Quantitative Financial Economics} 
(Wiley, Chichester, 1996);
J. C. Hull, {\it Options, Futures, and other Derivative Securities},
3rd Ed. (Prentice Hall, N.Y., 1997).

\bibitem{stan} R. N. Mantegna and H. E. Stanley, Nature {\bf 376}, 46 (1994).

\bibitem{bouchaud} For a brief review see J.-Ph. Bouchaud,
to be published (http://xxx.lanl.gov/archive/cond-mat/9806101).

\bibitem{bps}
P. Bak, M. Paczuski, and M. Shubik, Physica A {\bf 246}, 430 (1997).

\bibitem{ovch} A. A. Ovchinnikov, S. F. Timashev,
and A. A. Belyy, {\it Kinetics of Diffusion Controlled Chemical 
Processes} (Nova Science, Commack, 1990).

\bibitem{redner}
F. Leyvraz and S. Redner, Phys. Rev. Lett. {\bf 66}, 2168
(1991); E. Ben-Naim and S. Redner, J. Phys. A {\bf 25}, L575 (1992).

\bibitem{cornell}
S. Cornell and M. Droz, Phys. Rev. Lett. {\bf 70}, 3824 (1993);
Physica D {\bf 103}, 348 (1997), and references therein.

\bibitem{cardy} G. T. Barkema, M. J. Howard, and J. L. Cardy, 
Phys. Rev. E {\bf 53}, R2017 (1996).

\bibitem{krap} P. Krapivsky, Phys. Rev. E {\bf 51}, 4774 (1995).

\bibitem{note1} Close boundary conditions impose the conservation
of particle number in the system which is absent with open boundary
conditions. The effect of this is to reduce the validity of
Eq. (\ref{n_A_fluc}) to much shorter times, which in turn 
suppresses the market price fluctuations.

\bibitem{cahn} S. M. Allen and J. W. Cahn,
Acta. Metall. {\bf 27}, 1085 (1979).

\bibitem{note2} Copying in Model III introduces a weak correlation
among the $\eta_i$'s which does not affect our argument for
$\tau$ less than the lifetime of a particle.

\bibitem{note3} In the simulations we have measured $J$ directly
and observed that the ratio $\alpha/\alpha_c=8\gamma J/(\beta^2N)$
is slightly (up to $30\%$) bigger than one, but tends to one
as $N$ increases. We however do not have a satisfactory argument on
why $\alpha=\alpha_c$ is picked when $N$ is fixed.

\end{references}
\end{document}